\documentclass[twocolumn,secnumarabic, nobibnotes, aps, prb]{revtex4-1}
\usepackage{amssymb}
\usepackage{amsmath}
\usepackage{graphicx}
\newcommand{\mGaAs} {250-atom model I}
\newcommand{\mInN} {250-atom model II}

\begin{document}

\title{The properties of amorphous GaN }
\author{B. Cai}
\author{D. A. Drabold}
\affiliation{Dept. of Physics and Astronomy, Ohio University, Athens OH 45701, USA}
\date{\today }

\begin{abstract}
\textbf{Abstract}: In this paper, we present three amorphous GaN models obtained from the first principles simulation. We find that a chemically ordered continuous random network is the ideal structure for a-GaN. If we exclude the tail states, we predict a 3.0eV optical gap for 64-atom model and 2.3eV for 250-atom models. We observe a highly localized valence tail and a remarkably delocalized exponential conduction tail which we associate with different hybridization in the two tails. Based upon these results, we speculate on potential differences in \textit{n} and \textit{p} type doping. The structural origin of tail and defect states is discussed. The vibrational density of states and dielectric function are computed, and are consistent with experiment.
\end{abstract}

\maketitle

\affiliation{Dept. of Physics and Astronomy, Ohio University, Athens OH 45701, USA}

With a profound impact on lighting technology and other applications, crystalline GaN bas been the subject of vast inquiry\cite{RenGaN,PanGaN,PGaN}. However, lattice mismatch with substrates makes it difficult to grow. Recently, \textit{amorphous} GaN has become attractive due to its potential to solve the lattice match problem and its natural isotropy. A number of experiments have investigated the structural and optoelectronic properties of a-GaN\cite{aGaN1,Kod99,Kod02,OPTGaN,Expnohomo,Dielex}. In 1997, we proposed that a-GaN might find use as an electronic material\cite{Stum,Yuming}. 

The structure of a-GaN is controversial. Some experiments observe a large concentration of homopolar bonds\cite{exphomo}, contradicting other studies\cite{Expnohomo}. We did not observe N-N or Ga-Ga bonds in our computer models, but other calculations suggest a more disordered network\cite{theohomo}.  Doping of a-GaN is an important topic, but a full understanding of the intrinsic electronic features of undoped a-GaN, like the origin of defect states and tail states, is a necessary precursor. Naturally, the detailed properties depend upon the mode of growth of the material; our work is most relevant to least defective ``ideal" a-GaN.

It has been more than a decade since the first a-GaN model was proposed, a time during which many experiments have been carried out,  and simulation tools have experienced major developments. Therefore, additional calculations have been undertaken to generate atomistic models of a-GaN and to further explore its interesting properties.

In this paper, we propose atomistic a-GaN models formed \textit{via} \textit{ab initio} molecular dynamic simulation with a plane wave basis. The network topology is analyzed through radial and angular distribution functions, and structural statistics. We find that Ga and N atoms strongly prefer to be four-fold, and homopolar bonds are rare in the network. We also predict electronic properties and connect the electronic structure to the topology of the network. We show that the conduction edge has Urbach (exponential) form and is extraordinarily delocalized,  and the valence edge is very sharp with highly localized states. Doping is briefly discussed. Vibrational properties and dielectric functions are predicted.

\begin{table*}
\caption{The statistical distribution of the main structural components of three models.}
\label{table1}%
\begin{ruledtabular}
\begin{tabular}{ccccccccccc}
Model   &N-N  &Ga-Ga &$N_{3}$ &$N_{4}$ &$N_{5}$ &$Ga_{3}$ &$Ga_{4}$ &$Ga_{5}$ &$n_{N}$ &$n_{Ga}$\\
\hline
64-atom &0    &0     &3\%     &97\%    &0       &6\%      &91\%     &3\%      &3.97    &3.97 \\
\mGaAs  &0    &0     &11\%    &87\%    &2\%     &9\%      &89\%     &1\%      &3.91    &3.91\\
\mInN   &2\%  &0     &9\%     &89\%    &2\%     &11\%     &87\%     &2\%      &3.92    &3.90\\
\end{tabular}
\end{ruledtabular}
\end{table*}

All calculations in this work are performed with the Vienna Ab-Initio Simulation Package (VASP)\cite{VASP} based on density functional theory (DFT) within the local density approximation (LDA) and Vanderbilt's ultrasoft pseudopotentials. Three models are generated via computer alchemy: a 64-atom model (obtained from a 64-atom InN model of Ref. \onlinecite{aInN}), \mGaAs ~(obtained from 250-atom a-GaAs model of Ref. \onlinecite{aAsGa}) and \mInN ~(obtained from 250-atom a-InN model of Ref. \onlinecite{aInN}). The 64-atom model is annealed and equilibrated at 500K and \mGaAs~ is annealed and equilibrated at 300K. After zero-pressure relaxation, the density of the 64-atom model is $5.8g/cm^{3}$, 95\% of the crystalline GaN density; both 250-atom models have mass density around $5.6g/cm^{3}$, 92\% of the crystalline GaN density. The new 64-atom model shows a improved cohesive energy (lower by 0.16eV/atom) compared with the previous model\cite{Stum}. Where chemical order is concerned,  there are no homopolar bonds in the 64-atom model and \mGaAs, and only one N-N bond in the \mInN as one might expect for a partly ionic system. For coordination, we note that most atoms tend to be four-fold, suggesting that a-GaN retains the zinc-blende/wurtzite character of crystalline GaN,  in significant contrast to our early study\cite{Stum}. We list the structural properties in Table \ref{table1}. Since the 64-atom a-GaN was made from melt-and-quench method, while the both 250-atom models are generated only by a energetic relaxation, the little increase of over- and under-coordinated atoms in both 250-atom models could be the artifact of relaxation technique used. This result suggests that an ideal a-GaN is a chemically ordered continuous random network.

\begin{figure}
\begin{center}
\resizebox{1.0\columnwidth}{!}{
\includegraphics{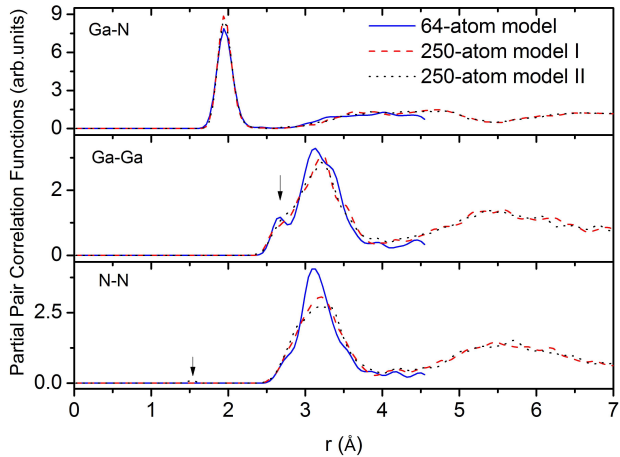}}
\end{center}
\caption{(Color online) Partial pair correlation functions of the three models (see text). In the Ga-Ga partial, the pre-peak/shoulders are marked by a black arrow. For the N-N partial, the black arrow marks the peak due to the only N-N bond in \mInN.}
\label{rdf}
\end{figure}

We plot the partial pair-correlation functions of the three models in Fig. \ref{rdf}. For the Ga-N partial, there exists a sharp first peak around 1.94\AA ~for all three models. The pre-peak/shoulders in Ga-Ga partial (marked by black arrow) indicate that there exists two local environment for Ga atoms and we will show that those two Ga sites are related to edge-sharing N tetrahedral structures. Due to the N-N bond, there is a small peak around $1.53\mathring{A}$~ in N-N partial of \mInN ~(marked by black arrow). Overall, the pair correlation functions of three models exhibit similar features and they are close to the results of Ref. \onlinecite{Yuming}.

Next, we analyze the angle distributions for Ga-N-Ga bonds and N-Ga-N bonds. The 250-atom models yield major peak positions close to $\theta_{T}=109.47^{\circ}$ for both Ga-N-Ga and N-Ga-N angle. For the 64-atom model,  the major peak positions are slightly off $\theta_{T}$, the mean value of of N-Ga-N and Ga-N-Ga angle, being $109.15^{\circ}$~and $108.65^{\circ}$ respectively. Thus, we conclude that a-GaN retains strong vestiges of its crystalline short-range order and tends to form a tetrahedral structure. We observe a pre-peak around $80^{\circ}$ for the 64-atom model (shoulders for the 250-atom model) in the Ga-N-Ga angle distribution which implies that there are two distinct sites for Ga atoms. After a detailed investigation, we find that the small angle is due to edge-sharing units with distorted angles (appearing as four-member-rings with Ga-N-Ga angle between $75^{\circ}$ and $95^{\circ}$). We will show that this kind of distortion is responsible for some of the electronic tail states.

\begin{figure}
\begin{center}
\resizebox{1.0\columnwidth}{!}{
\includegraphics{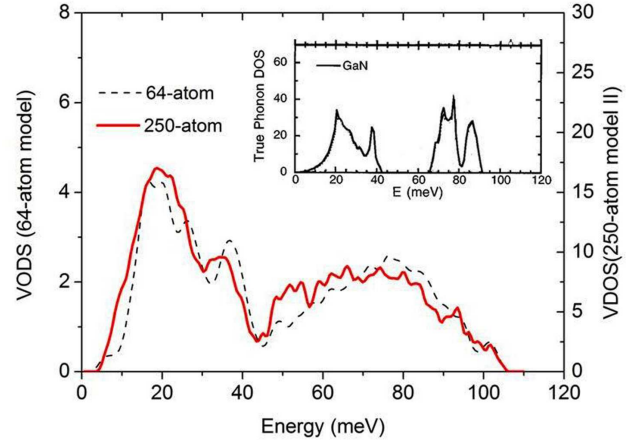}}
\end{center}
\caption{(Color online) Vibrational density of states of 64-atom model and \mInN. The eigenvalues were Gaussian broadened with a width of 1meV. The VODS of crystal wurtzite GaN is plotted as an inset from Ref. \onlinecite{VBxtGaN}.}
\label{VDOS}
\end{figure}

The vibrational properties of a-GaN are characterized through the vibrational density of states (VDOS). Starting with a thoroughly relaxed cell, the force constant and dynamical matrix is obtained from perturbing each atom in turn by $0.015 \AA$,  and computing forces on all atoms in the model for each perturbed conformation. The VDOS of the 64-atom model and \mInN~are reported in Fig. \ref{VDOS}. Both models show similar features. For comparison, we also plot the VODS of crystalline GaN from Ref. \onlinecite{VBxtGaN} as insert. Our results show that the amorphous VDOS retains some features of crystal VDOS such as the two peaks in the first band. However, we did not observe two distinguished peaks in the optical band\cite{VDOSaGaN}, and the gap between the acoustic band and the optical band fills in substantially. The results are quite consistent with a recent Raman study\cite{Expnohomo}.

\begin{figure}
\begin{center}
\resizebox{1.0\columnwidth}{!}{
\includegraphics{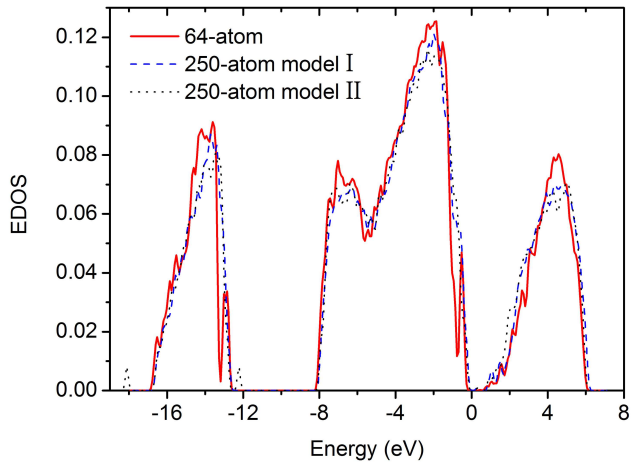}}
\end{center}
\caption{(Color online) Electronic density of states of 64-atom model(32 K-points are used), \mGaAs (8 K-points are used) and \mInN (8 K-points are used). The Fermi level is at 0 eV.}
\label{EDOS}
\end{figure}

\begin{figure}
\begin{center}
\resizebox{1.0\columnwidth}{!}{
\includegraphics{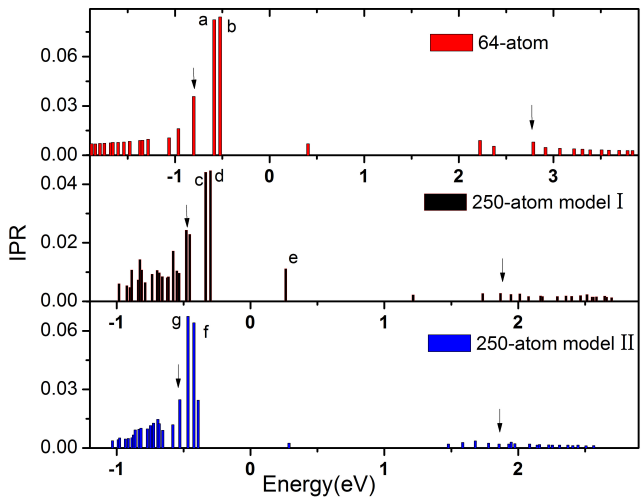}}
\end{center}
\caption{(Color online) Inverse participation ratio analysis for three models. Large IPR implies strong localization. For each model, the optical band gap are estimated by excluding the mid-gap and tail states (region between two black arrows in each plot). Note the highly delocalized conduction tail. The Fermi level is at 0eV.}
\label{IPR}
\end{figure}

We describe the electronic structure by analyzing the electronic density of states (EDOS), inverse participation ratio (IPR) of the individual states, and dielectric functions. Fig.\ref{EDOS} shows the EDOS of the three models. Overall, all three EDOS have similar character with slight differences in detail. The conduction band tail is \textit{dramatically} broader than the valence band tail. If we define the band gap as the difference between highest extended energy level of states in the valence band and lowest extended energy level of states in the conduction band (excluding the mid-gap and tail states), we estimate that the band gap to be about 3.0eV for the 64-atom model and 2.3eV for the 250-atom models. We indicate this in Fig.\ref{IPR} (the region between black arrows is taken as optical band gap). The band gaps obtained from our models are smaller than the experimental value 3.1eV\cite{OPTGaN}. This is primarily due to the LDA which is well-known to underestimate the band gap\cite{BinGST}. By fitting the conduction band tail to an exponential, we report the Urbach energy, $E_{u}\approx420meV$ for the 64-atom model and $E_{u}\approx490meV$ for the \mGaAs, comparable to the reported value ``several hundred meV" in Ref. \onlinecite{urbfilm}. In addition, for \mInN, there are defect states in the deep band region between -18eV and -12eV far below the Fermi level. These defects are due primarily to the N-N bond. Other than the defect states in deep band below -8eV, we did not observe any electronic signature of N-N bond around the optical gap.

\begin{figure*}[tbp]
\begin{center}
\resizebox{1.4\columnwidth}{!}{
\includegraphics{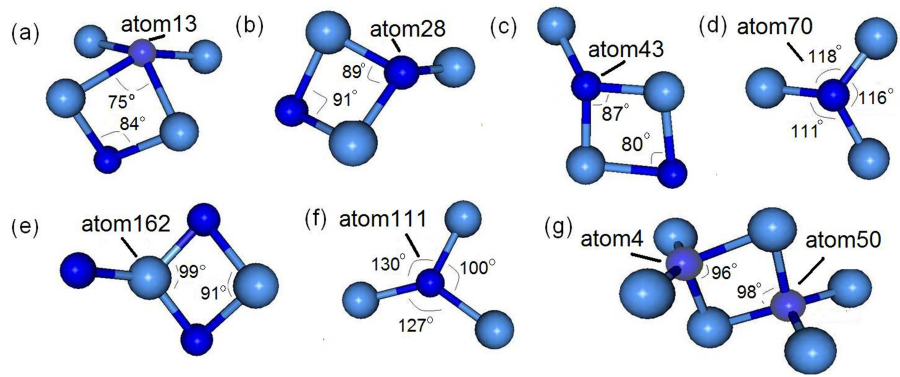}}
\end{center}
\caption{(Color online) Atomistic origin of electronic tail and gap states correlated with state \textit{a-g} indicated in Fig.\ref{IPR}. Dark(small) atom is N; light (large) atom is Ga. (a) atom-13 is associated with tail states \textit{a} and \textit{b}. (b) atom-28 is associated with tail state \textit{b}. (c) atom-43 is associated with tail states \textit{c} and \textit{d}. (d) atom-70 is associated with tail state \textit{d}. (e) atom-162 is associated with tail state \textit{e}. (f) atom-111 is associated with tail state \textit{f}. (g) atom-4 and atom-50 are associated with tail state \textit{g}.}
\label{unit}
\end{figure*}

To characterize the localization of the tail states around the gap, we performed an IPR analysis for all three models. The IPR measures the degree of localization given an electronic state\cite{anIPR}. For highly localized states, IPR=1; for extended states, IPR=$1/N$, where $N$ is the number of atoms. The results are plotted in Fig.\ref{IPR}. By projecting the EDOS onto different atomic orbitals, we find that the valence tail is built from N-p, Ga-p and Ga-d orbitals. This implies a high sensitivity to bond angle disorder, which is presumably the reason for high localization. The conduction tail is localized on Ga-s and N-s orbitals. Since the s-s interaction is \textit{only affected by bondlength}, the conduction tail states exhibit remarkably weak localization and the conduction tail is almost immune to angle disorder. This situation is somewhat similar to a-SiO$_{2}$, where there is also a large asymmetry on IPR between valence band and conduction band tails as discussed by Robertson\cite{Rob,ppsSiO}. To our knowledge, this effect has not been reported in nitrides. The asymmetry in width and localization of tail states suggests that \textit{n} and \textit{p} doping for a-GaN will be quite different\cite{ComparePL}. Due to the highly localized valence-band tail states, it will be more difficult to move the Fermi level toward the valence mobility edge, complex compensations may happen and mobility is likely to be poor. Thus, the p-type doping is expected to be relatively more difficult than n-type doping to obtain the same carrier concentration\cite{npdop}. 

To correlate electronic structure with topological units, we picked seven electronic states (\textit{a-g} in Fig.\ref{IPR}) with relatively high IPR and projected them onto individual atom sites. In Fig.\ref{unit}, we present the characteristic atomic sites associated with those tail states \textit{a-g}. In the 64-atom model, the tail states \textit{a} and \textit{b} are highly localized on atom-13 (Fig.\ref{unit}(a)) whose four neighbors are almost in the same plane,  with distorted Ga-N-Ga angle $75^{\circ}$; the state \textit{b} is also localized on atom-28(Fig.\ref{unit}(b)), the only three-fold N atom in the network which formed a small Ga-N-Ga angle near $89^{\circ}$. In \mGaAs, three-fold N atom-43 (Fig.\ref{unit}(c)), which formed a $87^{\circ}$ Ga-N-Ga angle, is strongly associated with eigenstate \textit{c} and \textit{d}; the tail state \textit{d} are also localized on atom-70 (Fig.\ref{unit}(d)), a three-fold N atom with all its neighbors almost in the same plane; moreover, the conduction-band tail state \textit{e} is mainly localized on atom-162 (Fig.\ref{unit}(e)), a three-fold Ga atom with disordered N-Ga-N angle. In \mInN, three-fold N atom-111 whose three neighbors are almost in the same plane (Fig.\ref{unit}(f)), contribute more to the valance-band tail state \textit{f}; two four-fold N atoms, atom-4 and atom-50, formed edge-shared tetrahedron with disordered Ga-N-Ga angle, are strongly associated with electronic state \textit{g}. Overall, atoms with distorted angle are associated with valence tail states. Finally, we briefly remark that, for the 64-atom model, the imaginary part of dielectric function $\epsilon(w)$ has a major peak position around 6.8eV for all three directions. This result is comparable to the experimental work reported in Ref. \onlinecite{Dielex}.

In conclusion, we created a-GaN atomistic models using state of the art methods. Most atoms in the network tend to be four-fold and form tetrahedral structures. We predict a 3.0eV optical band gaps for 64-atom model and 2.3eV for 250-atom model. We find an interesting and large asymmetry in localization between valence and conduction tail due to the different orbital interaction, which should yield quite distinct properties in \textit{n} and \textit{p} type doping. The atomistic origin of tail and defect states is discussed, and the disorder in bond angle is likely to introduce valence tail states, whereas the conduction tail is due primarily to bond length disorder. The vibrational density of states retains some qualitative features from the crystal,  and the dielectric functions shows a peak around 6.8eV, both of which are in agreement with experiment. Our work focuses primarily on ``ideal" GaN to establish a reference model. Ion bombarded samples are indeed likely to exhibit far more disorder\cite{exphomo,theohomo}.

We thank the National Science Foundation for support under grant DMR-09-03225. This work was supported in part by an allocation of computing time from the Ohio Supercomputer Center.


\begin{thebibliography}{9}
\bibitem{RenGaN} S. J. Pearton, F. Ren, A. P. Zhang, and K. P. Lee, Mater. Scie. and Engi. R: \textbf{R30}, 55-212 (2000).
\bibitem{PanGaN} Z. X. Zhang, X. J. Pan, T. Wang, E. Q. Xie and L. Jia, J. Alloy. Compd.\textbf{467}, 61-4 (2009).
\bibitem{PGaN} S. J. Pearton and F. Ren, Adv. Mater. \textbf{12}, No.21, (2000).
\bibitem{aGaN1} T. Hariu, T. Usuba, H. Adachi and Y. Shibata, Appl. Phys. Lett. \textbf{32}, 252 (1978).
\bibitem{Kod99} Z. Hassan, M. E. Kordesch, W. M. Jadwisienzak, H. J. Lozykowsky, W. Halverson and P. C. Colter, Microcrystalline and noncrystalline semiconductors-1998, \textbf{536}, 245-250 (1999).
\bibitem{Kod02} Z. Hassan, K. Ibrahim, M.E. Kordesch, W. Halverson and P. C. Cloter, Inter. J. of Mod. Phys. B, \textbf{16}, Nos.6 \& 7, 1086-1090 (2002).
\bibitem{OPTGaN} A. Al-Zouhbi and N. S. Al-din, Optical Review \textbf{15}, No.5, 251-254 (2008).
\bibitem{Expnohomo} A. Bittar, H. J. Trodahl, N. T. Kemp and A. Markwitz, Appl. Phys. Lett. \textbf{78}, 619 (2001).
\bibitem{Dielex} T. Kawashima, H. Yoshikawa, S. Adachi, S. Fuke and K. Ohtsuka, J. Appl. Phys. \textbf{82}, 3528 (1997).
\bibitem{Stum} P. Stumm and D. A. Drabold, Phys. Rev. Lett. \textbf{79}, 677 (1997).
\bibitem{Yuming} M. Yu and D. A. Drabold, Solid State Commun. \textbf{108}, 413 (1998).
\bibitem{exphomo} M. Ishimaru, Y. Zhang and W. J. Weber, J. of Appl. Phys. \textbf{106}, 053513 (2009).
\bibitem{theohomo} J. Nord, K. Nordlund and J. Keinonen, Phys. Rev. B. \textbf{68}, 184104 (2003).
\bibitem{VASP} G. Kresse and J. Furthmuller, VASP the GUIDE 2003, http://cms.mpi.univie.ac.at/vasp/.
\bibitem{aInN} B. Cai and D. A. Drabold, Phys. Rev. B. \textbf{79}, 195204 (2009).
\bibitem{aAsGa} N. Mousseau and G. T. Barkema, J. Phys.: Condens. Matter \textbf{16}, S5183 (2004).
\bibitem{VBxtGaN} J. C. Nipko, C. K. Loong, C. M. Balkas and R. F. Davis, Appl. Phys. Lett. \textbf{73}, 34 (1998).
\bibitem{VDOSaGaN} W. Pollard, J. of Non. Crsyt. Sol. \textbf{283}, 203-210 (2001).
\bibitem{BinGST} B. Cai, S. R. Elliott and D. A. Drabold, Appl. Phys. Lett. \textbf{97}, 191908 (2010).
\bibitem{urbfilm} S. Kobayashi, S. Nonomura, T. Ohmori, K. Abe, S. Hirata, T. Uno, T. Gotoh, S. Nitta and S. Kobayashi, Appl. Surf. Scie. \textbf{113-114}, 480-484 (1997).
\bibitem{anIPR} R. Atta-Fynn, P. Biswas and D. A. Drabold, Phys. Rev. B. \textbf{69}, 245204 (2004).
\bibitem{ppsSiO} F. Inam, J. P. Lewis and D. A. Drabold, Phys. Stat. Sol. a\textbf{207}, 599 (2010).
\bibitem{Rob} J. Robertson, J. of Non. Cryst. Sol. \textbf{354}, 2791-2795 (2008).
\bibitem{ComparePL} Y-H Kwon, S. K. Shee, G. H. Gainer, G. H. Park, S. J. Hwang and J. J. Song, Appl. Phys. Lett. \textbf{76}, 840 (2000).
\bibitem{npdop} C. G. Van de Walle, C. Stampfl and J. Neugebauer, J. of Non. Cryst. Sol. \textbf{189-190}, 505-510 (1998).

\end{thebibliography}
\end{document}